# Comment on «Hole transport and photoluminescence in Mg-doped InN»

## [J. Appl. Phys., 107, 113712 (2010)]

T.A. Komissarova* and S.V. Ivanov

*Ioffe Physical-Technical Institute, Polytekhnicheskaya 26, St.Petersburg 194021, Russia*

Analysis and discussion of the experimental results and conclusions reported in «Hole transport and photoluminescence in Mg-doped InN» [J. Appl. Phys., 107, 113712 (2010)] will be provided.

*) Author to whom correspondence should be addressed

electronic mail: komissarova@beam.ioffe.ru

Recently Miller et al. [1] have reported on hole transport in Mg-doped InN epilayers. Positive Seebeck coefficient and reduction of the Hall mobility have been considered as evidence of the existence of p-type conductivity in the bulk of the Mg-doped InN films. However, the proposed explanations of the experimental data do not seem to be sufficient to draw an unambiguous conclusion on observation of the p-type conductivity in the InN:Mg epilayers.

Hall effect measurements have revealed n-type conductivity for all the investigated Mg-doped InN films, while the Seebeck coefficient was positive for InN epilayers with the magnesium concentration [Mg] in the range of $3 \times 10^{17}$ - $2 \times 10^{19}$ cm$^{-3}$. The authors assert that the highly conductive n-type layer existing on the surface of all InN epilayers [2] prevents detection of p-type conductivity by Hall measurements in the Mg-doped InN films. Simultaneously they suppose that this n-type layer does not significantly influence the "hot probe" measurements and the positive Seebeck coefficient is related to the hole transport in the bulk of Mg-doped InN films. The contribution to the measured Seebeck coefficient from the n-type regions (film surface and interface) has been evaluated and taken into account using the model of parallel connection of n- and p-type regions in the Mg-doped InN films. Seebeck coefficient was calculated using the equation (1) in the frame of this model

$$S_{meas} = S_{bulk} \frac{\sigma_{bulk} d_{bulk}}{\sigma_{total} d_{total}} + S_n \frac{\sigma_n d_n}{\sigma_{total} d_{total}}, \qquad (1)$$

where $S_{meas}$ is the measured Seebeck coefficient, $S_{bulk}$, $\sigma_{bulk}$ and $d_{bulk}$ are the Seebeck coefficient, conductivity and thickness of the bulk of the InN film, $S_n$, $\sigma_n$ and $d_n$ are the corresponding parameters of the n-type regions, $\sigma_{total}$ and $d_{total}$ are the total conductivity and thickness of the InN film. To show that the p-type region can indeed dominate in the Seebeck effect measurements the authors have considered the following example: a 500 nm p-type film with the free hole concentration $p = 10^{18}$ cm$^{-3}$ and hole mobility $\mu_h = 30$ cm$^2$/Vs would have a Seebeck coefficient of approximately +600 µV/K, but adding the contribution of the 10-nm-thick n-type layer with conductivity 10 (Ω·cm)$^{-1}$ and $S_n = -40$ µV/K only reduce the observed Seebeck coefficient to $S_{meas} = 574$ µV/K. This value corresponds well to that measured in the Mg-doped InN films with $3\times10^{17} \leq [Mg] \leq 2\times10^{19}$ cm$^{-3}$. However, the authors did not consider the Hall coefficient $R_H$ in the same model of the parallel connection of n- and p-type layers, which can be calculated using equation (2)

$$R_H = \frac{1}{e} \frac{pd_p\mu_h^2 - nd_n\mu_e^2}{(pd_p\mu_h - nd_n\mu_e)^2}, \qquad (2)$$

where $n$, $p$ and $\mu_e$, $\mu_h$ are the electron and hole concentrations and mobilities of the n- and p-type layers, respectively; $d_n$ and $d_p$ are the corresponding thicknesses. After substitution of the electron and hole parameters from the above example into equation (2) one can obtain that the Hall coefficient must be also positive in this case, which contradicts the experimental values of the Hall coefficient being negative for all the samples. Therefore, simultaneous observation of the positive Seebeck and negative Hall coefficient in the particular Mg-doped InN films cannot be explained by the presence of the p-type bulk and n-type surface/interface layers connected in parallel. Hence, the main author's argument in favor of p-type conductivity in the bulk of Mg-doped InN films, based on the observation of the positive Seebeck effect, has not been proved properly.

The authors did not take into account that there exist several cases when the sign of the thermopower does not reflect the conductivity type, and the positive Seebeck coefficient can be observed for n-type material. Let us consider some possible reasons of the positive thermopower in the n-type InN films.

Firstly, InN is a degenerate semiconductor. The Seebeck coefficient can be determined by equation (3) within the Boltzman transport theory [3]:

$$S = \frac{\pi^2}{3}\frac{k}{e}kT\left[\frac{\partial \ln \sigma(E)}{\partial E}\right]_{E=E_F} \qquad (3)$$

The sign of thermopower is determined by the sign of the energy derivative of conductivity $\sigma(E)$. The conductivity is proportional to the density of states $\rho(E)$ and, hence, the sign of the Seebeck coefficient depends on the sign of energy derivative of $\rho(E)$ at Fermi energy [4]. The energy dependence of the density of states has complex structure in a degenerate semiconductor, as the impurity band overlaps energetically with the conduction band and the area with negative ∂ρ(E)/∂E arises. If the Fermi energy is located in this area then the positive thermopower can be observed in degenerate n-type material [5,6]. Mg-doping in the 3×10$^{17}$ - 2×10$^{19}$ cm$^{-3}$ range can lead to the shift of the Fermi level in the InN films to such energy region, causing observation of the positive Seebeck coefficient.

It should be also taken into account that there are two contributions to the thermopower effect: free carrier diffusion Sc and the phonon drag effect Sph. The thermopower related to the phonon drag effect has the same sign as Sc for normal electron-phonon scattering processes and the opposite sign for umklapp processes [3]. The total sign of the phonon drag contribution depends on whether the normal or umklapp scattering dominates. Therefore, if the phonon drag effect prevails and is reverse to the direction of electron diffusion, then the positive Seebeck coefficient is observed in the material with the electron conductivity.

One more possible reason for the positive Seebeck coefficient in the n-type InN films is their possible composite structure involving the InN semiconductor matrix and spontaneously formed metallic In nanoparticles. Evidences of the existence of In nanoparticles in some InN films grown in different laboratories have been obtained from optical measurements [7], high resolution transmission electron microscopy images [8], the low temperature features of the temperature dependence of resistivity [9,10], and the abnormal magnetic field dependence of the Hall coefficient [11]. The presence of metallic In inclusions in the InN films can lead to observation of the positive Seebeck coefficient in n-type InN. The thermopower of metallic indium is positive [12]. According to the expression for thermopower in composite materials, obtained in several papers [13,14], the thermopower of the composite material involving phases with the opposite Seebeck coefficients can be positive or negative depending on parameters of these phases. Due to Mg segregation effects in InN in a certain [Mg] range, Mg-doping can influence the distribution of the In inclusions in the InN films as well as their sizes and shapes, thus leading to the inversion of the sign of thermoelectric power. Additionally, Bergman and Fel [15] have calculated that the thermoelectric power in composites can be enhanced over that of the individual constituents. This fact could explain the large values of the Seebeck coefficient in the investigated InN films.

It is worth also noting that possible presence of In nanoparticles in the studied InN films can lead to the abnormal dependence of the Hall coefficient on magnetic field. In this case the electron mobility in the InN semiconductor matrix should be calculated from this dependence [11] rather than from Hall measurements at a single magnetic field. Bearing that in mind, the reduction of the Hall mobility in the Mg-doped InN films defined by Miller et al. from measurements at 0.3T [1] may not be necessarily caused by occurrence of p-type conductivity in the bulk of the InN films.

In conclusion, the author's explanation of the simultaneous observation of the positive Seebeck and negative Hall coefficients in some Mg-doped InN films by the existence of parallel electron (surface, interface) and hole (bulk) conductivity channels is not correct. Therefore the assertion about the existence of free holes in the bulk of the Mg-doped InN films still raises doubts. Furthermore, the authors did not consider alternative causes for the positive Seebeck coefficient in materials with purely electron conductivity, namely, the negative energy derivative of density-of-states at Fermi level in the degenerate semiconductor, the phonon drag effect, and the composite structure of material involving phases with opposite signs of the Seebeck coefficient.